\documentclass[traditabstract]{aa} % for the abstract without structuration
\usepackage{graphicx}
\usepackage{txfonts}
\usepackage{natbib}
\usepackage{amsmath}
\usepackage{ulem} 
\usepackage{lscape}

\usepackage{indentfirst}
\bibpunct[]{(}{)}{;}{a}{}{,}

\newcommand{\degs}{$^{\circ}$}

\usepackage{threeparttablex} % for "ThreePartTable" environment
\usepackage{booktabs}        % for well-spaced horizontal rules

\begin{document}  

   \title{Rotational spectroscopy of isotopic species of methyl mercaptan at millimeter and submillimeter 
          wavelengths: CH$_3$$^{34}$SH \thanks{The input and output files of the fit are available 
          as text files at CDS via anonymous ftp to cdsarc.u-strasbg.fr (130.79.128.5) or via 
          http://cdsweb.u-strasbg.fr/cgi-bin/qcat?J/A+A/.}$^,$\thanks{This manuscript is dedicated 
          to the memory of Li-Hong Xu who passed away at the final stage of writing of the manuscript.}}

   \author{Olena Zakharenko\inst{1}
           \and
           Frank Lewen\inst{1}
           \and
           Vadim V. Ilyushin\inst{2,3}
           \and 
           Holger S.~P. M{\"u}ller\inst{1}
           \and
           Stephan Schlemmer\inst{1}
           \and
           Eugene A. Alekseev\inst{2,3}
           \and
           Igor Krapivin\inst{2}
           \and
           Li-Hong Xu\inst{4}
           \and
           Ronald M. Lees\inst{4}
           \and
           Robin Garrod\inst{5}
           \and
           Arnaud Belloche\inst{6}
           \and
           Karl M. Menten\inst{6}
           }

   \institute{I.~Physikalisches Institut, Universit{\"a}t zu K{\"o}ln,
              Z{\"u}lpicher Str. 77, 50937 K{\"o}ln, Germany\\
              \email{zakharenko@ph1.uni-koeln.de, hspm@ph1.uni-koeln.de}
              \and
              Institute of Radio Astronomy of NASU, Mystetstv 4, 61002 Kharkiv, Ukraine 
              \and
              Quantum Radiophysics Department, V.~N. Karazin Kharkiv National University, Svobody Square 4, 61022 Kharkiv, Ukraine
              \and
              Department of Physics, University of New Brunswick, Saint John, NB, Canada
              \and
              Departments of Chemistry and Astronomy, The University of Virginia, Charlottesville, VA, USA
              \and
              Max-Planck-Institut f{\"u}r Radioastronomie, Auf dem H{\"u}gel 69, 53121~Bonn, Germany
              }

   \date{Received 29 March 2019 / Accepted 20 May 2019}

%%%%%%%%%%%%%%%%%%%%%%%%%%%%%%%%%%%%%%%%%%%%%%%%%%%%%%%%%%%%%%%%%%%%%%%%%%%%%%%%%%%%%%%%%
%%%%%%%%%%%%%%%%%%%%%%%%%%%%%%%%%%%%%%%%%%%%%%%%%%%%%%%%%%%%%%%%%%%%%%%%%%%%%%%%%%%%%%%%%

%%%%%%%%%%%%%%%%%%%%%%%%%%%%%%%%%%%%%%%%%%%%%%%%%%%%%%%%%%%%%%%%%%%%%%%%%%%%%%%%%%%%%%%%%
%%%%%%%%%%%%%%%%%%%%%%%%%%%%%%%%%%%%%%%%%%%%%%%%%%%%%%%%%%%%%%%%%%%%%%%%%%%%%%%%%%%%%%%%%

\abstract{Methyl mercaptan (CH$_3$SH) is an important sulfur-bearing species in
 the interstellar medium, terrestrial environment, and potentially in planetary atmospheres.
 The aim of the present study is to provide accurate spectroscopic parameters for the most 
 abundant minor isotopolog CH$_3$$^{34}$SH to support radio astronomical observations 
 at millimeter and submillimeter wavelengths. The rotational spectrum of CH$_3$$^{34}$SH, 
 which is complicated by the large-amplitude internal rotation of the CH$_3$ group 
 versus the $^{34}$SH frame, was investigated in the 49$-$510~GHz and 1.1$-$1.5~THz 
 frequency ranges in natural isotopic abundance. 
 The analysis of the spectrum was performed up to the second excited torsional state, 
 and the obtained data were modeled with the RAM36 program. A fit within experimental accuracy was 
 obtained with a RAM Hamiltonian model that uses 72 parameters. Predictions based on this fit are 
 used to search for CH$_3$$^{34}$SH with the Atacama Large Millimeter/submillimeter Array 
 (ALMA) toward the hot molecular core Sgr~B2(N2), but blends with emission lines of other species 
 prevent its firm identification in this source.}

\keywords{Methods: laboratory: molecular -- Techniques: spectroscopic -- ISM: molecules -- Astrochemistry -- ISM: abundances -- Radio lines: ISM}

\authorrunning{O. Zakharenko et al.}
\titlerunning{Laboratory spectroscopic study of CH$_3$$^{34}$SH}

\maketitle
\hyphenation{For-schungs-ge-mein-schaft}

%%%%%%%%%%%%%%%%%%%%%%%%%%%%%%%%%%%%%%%%%%%%%%%%%%%%%%%%%%%%%%%%%%%%%%%%%%%%%%%%%%%%%%%%%
%%%%%%%%%%%%%%%%%%%%%%%%%%%%%%%%%%%%%%%%%%%%%%%%%%%%%%%%%%%%%%%%%%%%%%%%%%%%%%%%%%%%%%%%%
\section{Introduction}
\label{intro}
%%%%%%%%%%%%%%%%%%%%%%%%%%%%%%%%%%%%%%%%%%%%%%%%%%%%%%%%%%%%%%%%%%%%%%%%%%%%%%%%%%%%%%%%%
%%%%%%%%%%%%%%%%%%%%%%%%%%%%%%%%%%%%%%%%%%%%%%%%%%%%%%%%%%%%%%%%%%%%%%%%%%%%%%%%%%%%%%%%%

Sulfur (S) is of importance for astrophysics as one of the constituents of interstellar dust. 
The studies of the interstellar grains, extracted from the interplanetary dust particles 
embedded in meteorites, show that sulfur is often included into interstellar silicates 
\citep{Bradley:1999}. 
Astronomical silicates are one of the fundamental building blocks from which the solar
system is assumed to be formed. In addition, sulfur is considered to be one of the tools to
study the evolutionary stages of massive stars \citep{Kahane:1988}.

The sulfur-bearing molecule methyl mercaptan, CH$_3$SH, plays an important role in interstellar chemistry and 
potentially in planetary atmospheres \citep{Vance:2011}. It is abundant in some astronomical sources and 
was first detected in Sgr B2 \citep{Turner:1975}. The molecule was observed later toward the high-mass star-forming 
region G327.3$-$0.6 \citep{Gibb:2000}. Recently, methyl mercaptan was observed 
in molecular line surveys carried out with the Atacama Millimeter/submillimeter Array (ALMA) 
towards Sgr B2(N2) \citep{Holger2016,2016A&A...587A..91B} at levels that make detection of its $^{34}$S 
isotopic species promising. The terrestrial abundance of the second-most abundant stable sulfur 
isotope $^{34}$S is 4.21\%. Investigation of the $^{34}$S isotopolog and its detection in 
the interstellar medium will provide additional information on the $^{32}$S/$^{34}$S abundance 
ratio, which is of interest for investigation of stellar nucleosynthesis and thus Galactic chemical evolution.
With the increased sensitivity and resolution of the ALMA telescope, a large number of new 
unknown lines is observed compared to the past. A considerable portion of the unknown lines 
belongs to high rotational or vibrational (including torsional) states of known molecules, 
as well as to their isotopologs, with the CH$_3$$^{34}$SH being a potential contributor.

The rotational spectrum of methyl mercaptan was investigated in the centimeter
\citep{Kojima:1957,Kojima:1960}, millimeter \citep{Bettens:1999,Lees:1980,Sastry:1986,Xu:2012}, 
and infrared wavelength regimes \citep{Lees:2018}. Recently, the CH$_3$$^{32}$SD isotopolog 
was studied in the 150$-$510 GHz frequency range \citep{Zakharenko:2019SD}. 
This study extended the previous investigation of the CH$_3$$^{32}$SD spectrum performed 
by \citet{TSUNEKAWA:1989}. Concerning the $^{34}$S isotopolog of methyl mercaptan, 
only one transition, $1_{01}\leftarrow0_{00}$, was observed \citep{Solimene:1955}. 
One additional series of the $^{34}$S isotopic species of the molecule was found 
around 31~GHz by \citet{Kojima:1960}, but assignments were not discussed. 
We report a global modeling of the rotational transitions of CH$_3$$^{34}$SH in the ground, 
the first, and the second excited torsional states. The spectrum is complicated by the 
large-amplitude internal rotation of the methyl group CH$_3$ with respect to the $^{34}$SH frame. 
Quantum chemical calculations were performed to provide initial estimates of the rotational 
constants and determine the geometry of CH$_3$$^{34}$SH. 
The aim of the present study is to obtain accurate spectroscopic parameters of the 
$^{34}$S isotopolog of methyl mercaptan to support astronomical observations by radio 
telescope arrays, in particular at millimeter and submillimeter wavelengths.
The remainder of the manuscript is laid out as follows. Sections~\ref{calc} and \ref{exptl} 
provide details on our quantum chemical calculations and on our laboratory measurements, 
respectively. The laboratory spectroscopic analysis and discussion are given in 
Section~\ref{analysis-discuss}. Sections~\ref{astro-obs} and \ref{astro-results} 
describe our astronomical observations and the results of our search for CH$_3$$^{34}$SH, 
respectively, while Section~\ref{conclusion} details the conclusions of our investigations.

%%%%%%%%%%%%%%%%%%%%%%%%%%%%%%%%%%%%%%%%%%%%%%%%%%%%%%%%%%%%%%%%%%%%%%%%%%%%%%%%%%%%%%%%%
%%%%%%%%%%%%%%%%%%%%%%%%%%%%%%%%%%%%%%%%%%%%%%%%%%%%%%%%%%%%%%%%%%%%%%%%%%%%%%%%%%%%%%%%%
\section{Quantum chemical calculations}
\label{calc}
%%%%%%%%%%%%%%%%%%%%%%%%%%%%%%%%%%%%%%%%%%%%%%%%%%%%%%%%%%%%%%%%%%%%%%%%%%%%%%%%%%%%%%%%%
%%%%%%%%%%%%%%%%%%%%%%%%%%%%%%%%%%%%%%%%%%%%%%%%%%%%%%%%%%%%%%%%%%%%%%%%%%%%%%%%%%%%%%%%%

\begin{figure}[h]
\centering
  \includegraphics[height=5cm]{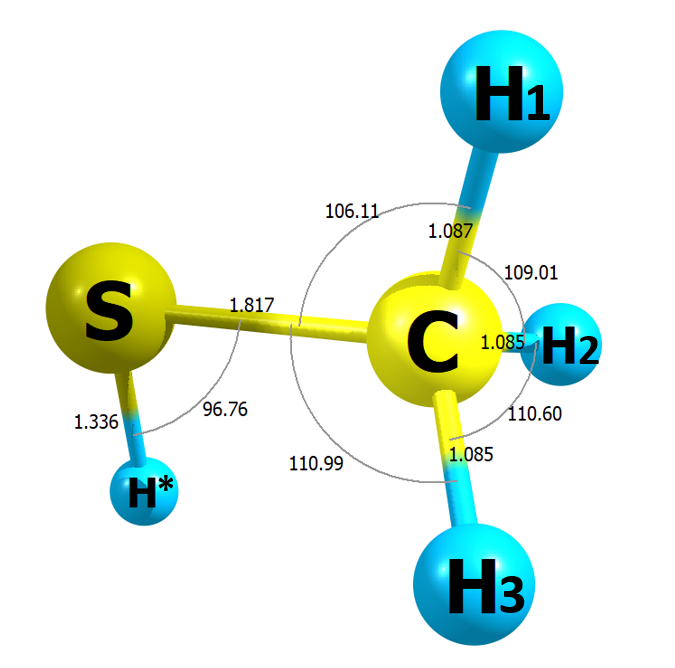}
  \caption{Schematic view of $^{34}$S methyl mercaptan and equilibrium geometry calculated 
    at MP2/aug-cc-pVTZ level of theory. The structural parameters are shown in angstroms and degrees.}  
  \label{fgr:example1}
\end{figure}

All calculations were performed using the Gaussian 09 (G09), Revision A.03, 
software package \citep{g16}. The molecular parameters and the harmonic force field of methyl 
mercaptan were evaluated using the M{\o}ller-Plesset second-order theory (MP2) 
\citep{PhysRev.46.618} with aug-cc-pVTZ basic set. The structural parameters were 
obtained in order to provide an accurate equilibrium structure. The schematic view 
and the main structural parameters are shown in Fig.~\ref{fgr:example1}.   

%%%%%%%%%%%%%%%%%%%%%%%%%%%%%%%%%%%%%%%%%%%%%%%%%%%%%%%%%%%%%%%%%%%%%%%%%%%%%%%%%%%%%%%%%
%%%%%%%%%%%%%%%%%%%%%%%%%%%%%%%%%%%%%%%%%%%%%%%%%%%%%%%%%%%%%%%%%%%%%%%%%%%%%%%%%%%%%%%%%

%%%%%%%%%%%%%%%%%%%%%%%%%%%%%%%%%%%%%%%%%%%%%%%%%%%%%%%%%%%%%%%%%%%%%%%%%%%%%%%%%%%%%%%%%
%%%%%%%%%%%%%%%%%%%%%%%%%%%%%%%%%%%%%%%%%%%%%%%%%%%%%%%%%%%%%%%%%%%%%%%%%%%%%%%%%%%%%%%%%
\section{Experimental details}
\label{exptl}
%%%%%%%%%%%%%%%%%%%%%%%%%%%%%%%%%%%%%%%%%%%%%%%%%%%%%%%%%%%%%%%%%%%%%%%%%%%%%%%%%%%%%%%%%
%%%%%%%%%%%%%%%%%%%%%%%%%%%%%%%%%%%%%%%%%%%%%%%%%%%%%%%%%%%%%%%%%%%%%%%%%%%%%%%%%%%%%%%%%

Measurements in Cologne were done in the frequency ranges of 155$-$510~GHz and 1.1$-$1.5~THz
using the Cologne mm/submm wave and THz spectrometers. An Agilent E8257D synthesizer, referenced 
to a rubidium standard, together with VDI (Virginia Diodes, Inc.) Amplified Multiplier Chains 
were employed as frequency sources. The mm/submm output radiation was guided through the 
5-m double-pass glass cell (the THz radiation to the 7-m glass cell) of 10-cm in diameter 
and then to the detectors. We used Schottky diode detectors to detect output frequencies 
in the 155$-$510~GHz frequency region and a helium-cooled bolometer (QMC Instruments Ltd.)
in the 1.1$-$1.5~THz frequency region. The measurements were carried out at room temperature 
and at a pressure of 20$-$40~$\mu$bar. The input frequency was modulated at 47.8~kHz 
in the 155$-$510~GHz range and at 16.7~kHz in the terahertz range. The modulation amplitude 
and frequency steps were adjusted to optimize the S/N ratio. The output signal from 
the detector was detected by a lock-in amplifier in 2\textit{${f}$} mode to give
second-derivative spectra, with a time constant of 20 or 50~ms. Detailed descriptions 
of the spectrometers may be found in \citet{Bossa:2014} and \citet{Xu:2012}. 
Methyl mercaptan (purity $\ge 98.0\%$) was purchased from Sigma Aldrich and was used 
without further purification. The $^{34}$S isotopic species of CH$_3$SH was measured 
in natural abundance.

Measurements in Kharkiv were done in the frequency range of 49$-$150~GHz using
the automated spectrometer of the Institute of Radio Astronomy of NASU (\citet{Alekseev:2012}).
The synthesis of the frequencies in the millimeter wave range was carried out by a two-step 
frequency multiplication of a reference synthesizer in two phase-lock-loop (PLL) stages. 
The reference synthesizer is a computer-controlled direct digital synthesizer (DDS AD9851), 
whose output is upconverted into the 385$-$430~MHz frequency range. At the first multiplication 
stage a klystron operating in the 3.4$-$5.2~GHz frequency range with a narrowband (1~kHz) PLL 
system was used. At the second multiplication stage, an Istok backward wave oscillator (BWO) was 
locked to a harmonic of the klystron. A set of BWOs was used to cover the frequency 
range from 49 to 149~GHz. The input frequency was modulated at 11.16~kHz and the output signal 
from the detector was detected by a lock-in amplifier in 1$f$ mode to give 
first-derivative spectra. The measurements were carried out at room temperature and 
at a pressure of 10$-$20~$\mu$bar. The uncertainty of the measurements was estimated to be 
10 kHz for a relatively strong isolated line (S/N > 10), 30 kHz for weak lines (2 < S/N ratio < 10),
and 100 kHz for very weak lines (S/N < 2). Methyl mercaptan was synthesized from a 21\% water 
solution of sodium thiomethoxide CH$_3$SNa (purchased from Sigma Aldrich and used without further 
purification). The $^{34}$S isotopic species of CH$_3$SH was measured in natural abundance.

%%%%%%%%%%%%%%%%%%%%%%%%%%%%%%%%%%%%%%%%%%%%%%%%%%%%%%%%%%%%%%%%%%%%%%%%%%%%%%%%%%%%%%%%%
%%%%%%%%%%%%%%%%%%%%%%%%%%%%%%%%%%%%%%%%%%%%%%%%%%%%%%%%%%%%%%%%%%%%%%%%%%%%%%%%%%%%%%%%%
\section{Laboratory spectroscopic analysis and discussion} 
\label{analysis-discuss}
%%%%%%%%%%%%%%%%%%%%%%%%%%%%%%%%%%%%%%%%%%%%%%%%%%%%%%%%%%%%%%%%%%%%%%%%%%%%%%%%%%%%%%%%%
%%%%%%%%%%%%%%%%%%%%%%%%%%%%%%%%%%%%%%%%%%%%%%%%%%%%%%%%%%%%%%%%%%%%%%%%%%%%%%%%%%%%%%%%%
\begin{figure*}
 \centering
 \includegraphics[height=10cm]{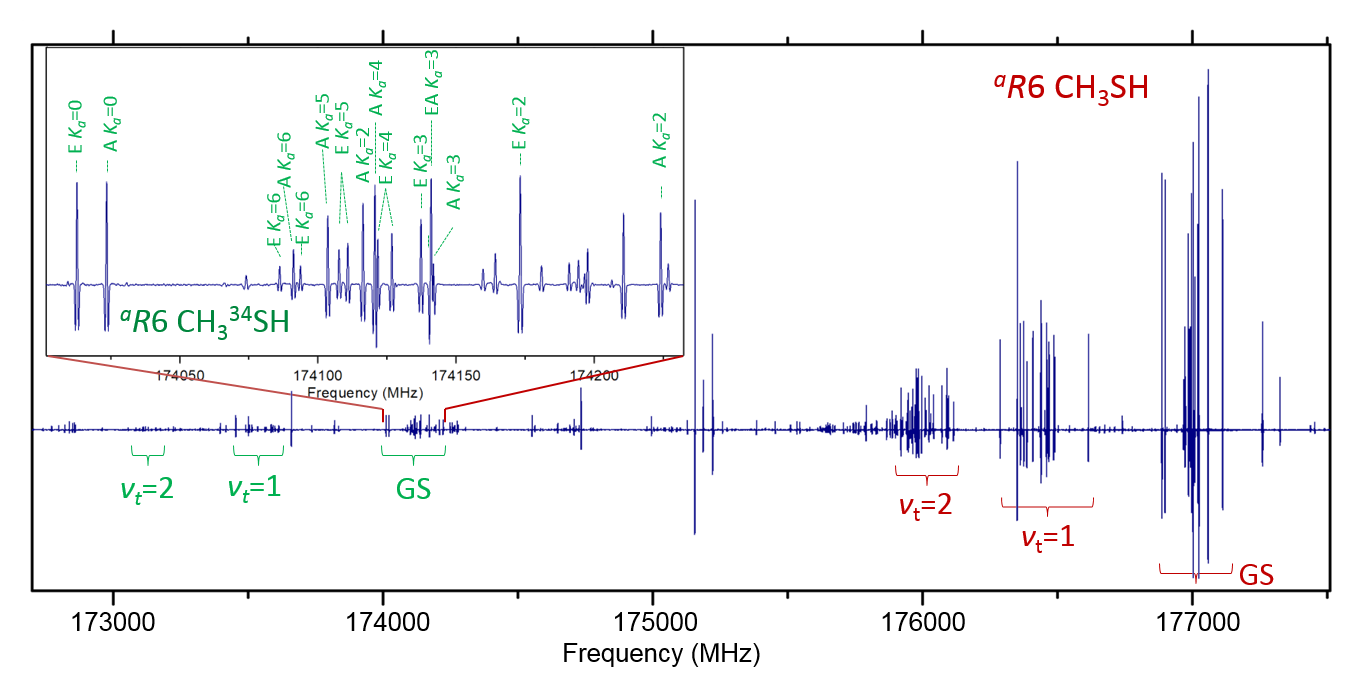}
 \caption{Partial methyl mercaptan spectrum around 175~GHz. The $^aR6$ rotational transitions 
  in the ground as well as the first two excited torsional states of CH$_3$$^{32}$SH are shown 
  on the right and those of CH$_3$$^{34}$SH on the left. The inset contains symmetry (A or E) 
  and $K_a$ labels for lines of the ground state of CH$_3$$^{34}$SH.}
 \label{fgr:example2col}
\end{figure*}
%%%%%%%%%%%%%%%%%%%%%%%%%%%%%%%%%%%%%%%%%%%%%%%%%%%%%%%%%%%%%%%%%%%%%%%%%%%%%%%%%%%%%%%%%
%%%%%%%%%%%%%%%%%%%%%%%%%%%%%%%%%%%%%%%%%%%%%%%%%%%%%%%%%%%%%%%%%%%%%%%%%%%%%%%%%%%%%%%%% 

Methyl mercaptan has two dipole moment components, $\mu_a=1.312$~D and $\mu_b=0.758$~D 
\citep{TSUNEKAWA:1989}, thus, both $a$-type and $b$-type transitions are observed. 
The $a$ axis lies almost parallel to the S$-$C bond at an angle of 1.67\degs , and the 
H$^{*}-$S$-$C$-$H{\tiny2} chain lies in the $ab$ plane with the H$^{*}-$S bond at an angle 
of 8.43\degs \ to the $b$ axis, see Fig.~\ref{fgr:example1}.
The $\rho$ axis method (RAM) \citep{Kirtman:1962,Lees:1968,HOUGEN:1994}, implemented 
in the RAM36 code \citep{Ilyushin:2010}, was used to perform the analyses of the high 
resolution millimeter-wave and terahertz spectra of methyl mercaptan. 
The initial predictions were based on a set of the rotational constants obtained from 
quantum chemical calculations. With the assumption that the $^{34}$S isotopic substitution 
does not considerably alter the molecular structure, the values of the quartic centrifugal 
distortion parameters and main internal rotation parameters (such as the internal rotation 
barrier $V_3$, the coupling parameter $\rho$, and the internal rotation constant $F$) 
were fixed at the corresponding values of the CH$_3$$^{32}$SH parent molecule \citep{Xu:2012}.

We started our analysis from the search of the series of intense $a$-type $R$-branch 
transitions with low $K_a$ quantum numbers. The part of the spectrum of methyl mercaptan 
illustrated in Fig.~\ref{fgr:example2col} shows the relative intensities of the series of 
$^aR6$ rotational transitions of CH$_3$$^{34}$SH compared to the main isotopolog.  
The predicted lines were not in the immediate vicinity of the observed lines. 
Nevertheless, the assignments of the ground state lines of CH$_3$$^{34}$SH were made due 
to the similarity in the spectral patterns of the $^{34}$S and the main isotopic species. 
The assigned transitions were fit, and further analyses proceeded by increasing 
the $J$ and $K_a$ quantum numbers. The process of assignments comprises 
gradual adding of newly measured transitions to the dataset and numerous cycles of 
improving of the parameter set. A large number of $Q$-, $P$-, and $R$-branch $b$-type 
transitions were assigned subsequently, which provided a large dataset for the detailed 
modeling of the rotational spectrum.

%%%%%%%%%%%%%%%%%%%%%%%%%%%%%%%%%%%%%%%%%%%%%%%%%%%%%%%%%%%%%%%%%%%%%
%%%%%%%%%%%%%%%%%%%%%%%%%%%%%%%%%%%%%%%%%%%%%%%%%%%%%%%%%%%%%%%%%%%%%
%%%%%%    Table 1    %%%%%%%%%%%%%%%%%%%%%%%%%%%%%%%%%%%%%%%%%%%%%%%%
%%%%%%%%%%%%%%%%%%%%%%%%%%%%%%%%%%%%%%%%%%%%%%%%%%%%%%%%%%%%%%%%%%%%%
%%%%%%%%%%%%%%%%%%%%%%%%%%%%%%%%%%%%%%%%%%%%%%%%%%%%%%%%%%%%%%%%%%%%%

\begin{table}[t]

\small
  \caption{\ Total number of transitions and other statistical information for the CH$_3$$^{34}$SH data set.}
  \label{tbl:statisticInf}
  \begin{tabular*}{0.48\textwidth}{@{\extracolsep{\fill}}rrrrr}
    \hline
    $\varv_t$, sym$^{a}$ & $N^{b}$ & $K_{a,\rm {max}}$$^{c}$ & $J_{\rm max}$$^{d}$ & rms$^{e}$ \\  
    \hline

    0, A & 993 & 16 & 55 & 41.0 \\
    0, E & 918 & 15 & 55 & 39.5 \\
    1, A & 504 & 13 & 45 & 39.1 \\
    1, E & 501 & 12 & 47 & 27.9 \\
    2, A & 235 &  7 & 38 & 32.2 \\
    2, E & 318 &  8 & 38 & 63.6 \\
   
    \hline
  \end{tabular*}
\tablefoot{
 $^{a}$ Torsional level and symmetry label (A or E) of lower and upper states of rotational transition. 
 $^{b}$ Number of rotational transitions in a given category.
 $^{c}$ Maximum value of $K_a$ quantum number in a given category.
 $^{d}$ Maximum value of $J$ quantum number in a given category. 
 $^{e}$ Root mean square (rms) deviation for corresponding group of data in kHz.
}
\end{table}
%%%%%%%%%%%%%%%%%%%%%%%%%%%%%%%%%%%%%%%%%%%%%%%%%%%%%%%%%%%%%%%%%%%%%%%%%%%%%%%%%%%%%%%%%
%%%%%%%%%%%%%%%%%%%%%%%%%%%%%%%%%%%%%%%%%%%%%%%%%%%%%%%%%%%%%%%%%%%%%%%%%%%%%%%%%%%%%%%%%

\begin{figure*}
 \centering
 \includegraphics[height=12cm]{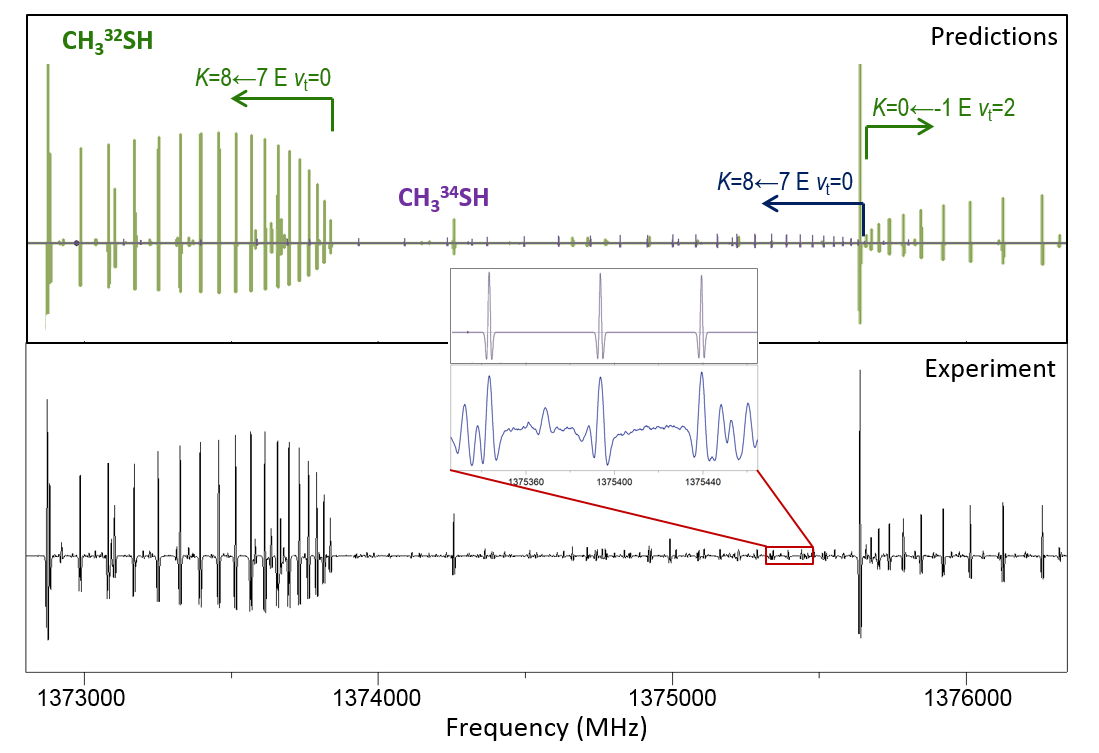}
 \caption{Partial methyl mercaptan spectrum around 1375~GHz. Series of $Q$ branch rotational transitions 
  in the THz frequency region are shown ($K=8\leftarrow7,~E,~\varv_t=0$ and $K=0\leftarrow -1,~E,~\varv_t=2$ 
 of CH$_3$$^{32}$SH, and $K=8\leftarrow7,~E,~\varv_t=0$ of CH$_3$$^{34}$SH).}
 \label{fgr:BildTHzSpektrum}
\end{figure*}

The analyses of the first and second excited states of the methyl torsion were carried 
out in a similar manner. The predictions for the excited torsional states were calculated based 
on the set of the ground state parameters of CH$_3$$^{34}$SH. Finally, a number of measured 
transitions in the frequency range 1.1$-$1.5 THz were added to the dataset. 
These data did not pose any problem for the fitting process.  
Figure~\ref{fgr:BildTHzSpektrum} shows an example of the agreement achieved between predicted 
and experimental spectra for methyl mercaptan in the THz region, with the three $Q$-branch series 
again illustrating the relative intensity ratio between the main and $^{34}$S isotopic species.

%%%%%%%%%%%%%%%%%%%%%%%%%%%%%%%%%%%%%%%%%%%%%%%%%%%%%%%%%%%%%%%%%%%%%%%%%%%%%%%%%%%%%%%%%
%%%%%%%%%%%%%%%%%%%%%%%%%%%%%%%%%%%%%%%%%%%%%%%%%%%%%%%%%%%%%%%%%%%%%%%%%%%%%%%%%%%%%%%%%
%%%%%%%%%%%%%%%%%%%%%%%%%%%%%%%%%%%%%%%%%%%%%%%%%%%%%%%%%%%%%%%%%%%%%%%%%%%%%%%%%%%%%%%%%
%%%%%%%%%%%%%%%%%%%%%%%%%%%%%%%%%%%%%%%%%%%%%%%%%%%%%%%%%%%%%%%%%%%%%%%%%%%%%%%%%%%%%%%%%

In total, 3469 rotational transitions were measured for the ground, the first, and the second  
excited torsional states of CH$_3$$^{34}$SH, which correspond to 3069 fit line frequencies 
due to blending. Some statistical information on the final fit is presented in 
Table~\ref{tbl:statisticInf}. The estimated uncertainties for measured line frequencies are 
in the range from 10 to 100 kHz depending on the frequency range and the observed S/N ratio. 
The upper limits of the $J$ and  $K_a$ quantum numbers naturally decrease with 
energy progression from the ground, over the first, to the second excited torsional states. 
The full dataset was fit using 72 parameters with an overall weighted root mean square (rms) 
deviation of 0.83. 
The input and output files of the final fit are available on the Cologne Database for 
Molecular Spectroscopy (CDMS) \citep{ENDRES201695}. The 72 parameters from the global fit 
of the CH$_3$$^{34}$SH spectrum are listed in Table~\ref{tbl:ParametersTable}. 
It should be noted that the final fit converged properly: the relative change in 
the weighted rms deviation of the fit at the last iteration was about $10^{-7}$; 
the corrections to the parameter values generated at the last iteration are less than $10^{-4}$ 
of the calculated parameter confidence intervals; the changes generated at the last iteration
in the calculated frequencies are less than 1~kHz. Nevertheless, the analysis of the singular 
value decomposition reveals a group of highly correlated parameters, namely 
$\rho_{mK}$, $\rho_{KK}$, $\rho_{mm}$, $F_{mm}$, $F_{mK}$, $F_{KK}$, $\rho_{m}$, $F_{K}$. 
Unfortunately, elimination of any of these correlated parameters leads to a significant worsening 
of the overall weighted rms deviation of the fit. The problem with the correlations listed above 
may be caused by small rotation-torsion-vibration interactions in the spectra, 
such as those discussed in \citet{Xu:2012} in connection with the analysis of the main isotopolog 
of methyl mercaptan. Similar perturbations occur in the spectrum of the $^{34}$S isotopolog and 
increased rms deviations for the $E$ type; $\varv_t=2$ group of transitions in comparison with 
other groups is a manifestation of such perturbations. 
The perturbations from the low-lying, small amplitude vibrations propagating through torsion-rotation 
interaction down to the low-lying torsional states require inclusion of additional terms that lead 
to increased correlations among parameters. An additional manifestation of such perturbations 
in the parameter set is the same order of magnitude of the $V_6$ and $V_9$ terms 
in the expansion of the torsional potential function.

We decided to compare our current results with the parameters of the main isotopolog \citep{Xu:2012} 
and of the deuterated methyl mercaptan \citep{Zakharenko:2019SD} in order to have a more detailed 
picture of what is happening with the Hamiltonian model of CH$_3$SH upon isotopic substitution. 
All three fits employ slightly different sets of high-order torsion-rotational parameters, 
in particular due to differences in the datasets. Therefore, we limit ourselves to the comparison 
of parameters of only up to the fourth order in Table~\ref{tbl:ParametersComparison}. 
Compared to the H/D substitution, the $^{34}$S isotopic substitution does not lead to large 
changes either in the rotational constants or in the main internal rotational parameters 
$V_3$, $\rho$, and $F$ (the differences for the latter are about 0.1 percent and less). 
Furthermore, many torsion-rotation distortion parameters of the fourth order agree between 
the $^{34}$S and $^{32}$S isotopologs. 

Nevertheless, we observe changes in sign for two parameters, namely $\Delta_{JK}$ and $D_{abJ}$, 
for the $^{34}$S isotopolog compared to the $^{32}$S isotopolog. At the same time, the sign of the 
$\Delta_{JK}$ parameter in the $^{34}$S isotopolog coincides with the sign of the corresponding 
parameter in the SD isotopolog. Although these discrepancies may be caused by the difference 
in the datasets, it is also possible that the agreement in the sign of the $\Delta_{JK}$ parameter 
between the $^{34}$SH and $^{32}$SD isotopologs speaks in favor of additional correlation issues 
in the CH$_3$$^{32}$SH parameter set, where the number of fourth order parameters exceeds by two 
the number of allowed parameters for this order, as predicted by the reduction scheme proposed by 
\citet{NAKAGAWA:1987}.

\section{Observations}
\label{astro-obs}

We use the Exploring Molecular complexity with ALMA (EMoCA) spectral line survey performed 
toward the high-mass star-forming region Sagittarius~B2(N) with ALMA to search for the $^{34}$S 
isotopolog of methyl mercaptan. The observations, data reduction, and method used to identify 
the detected lines and derive column densities were described in detail in \citet{2016A&A...587A..91B}. 
In short, the survey has a median angular resolution of 1.6$\arcsec$. It was done with 
five frequency tunings that fully cover the frequency range from 84.1~GHz to 114.4~GHz 
with a spectral resolution of 488.3~kHz (1.7 to 1.3~km~s$^{-1}$). The phase center was set 
at ($\alpha, \delta$)$_{\rm J2000}= (17^{\rm h}47^{\rm m}19{\fs}87, -28^\circ22'16{\farcs}0$). 
We focus on the peak position of the hot molecular core Sgr~B2(N2) at 
($\alpha, \delta$)$_{\rm J2000}= (17^{\rm h}47^{\rm m}19{\fs}86$, $-28^\circ22\arcmin13{\farcs}4$) 
in the present work.

\section{Astronomical results}
\label{astro-results}

\begin{figure}
\centerline{\resizebox{0.98\hsize}{!}{\includegraphics[angle=0]{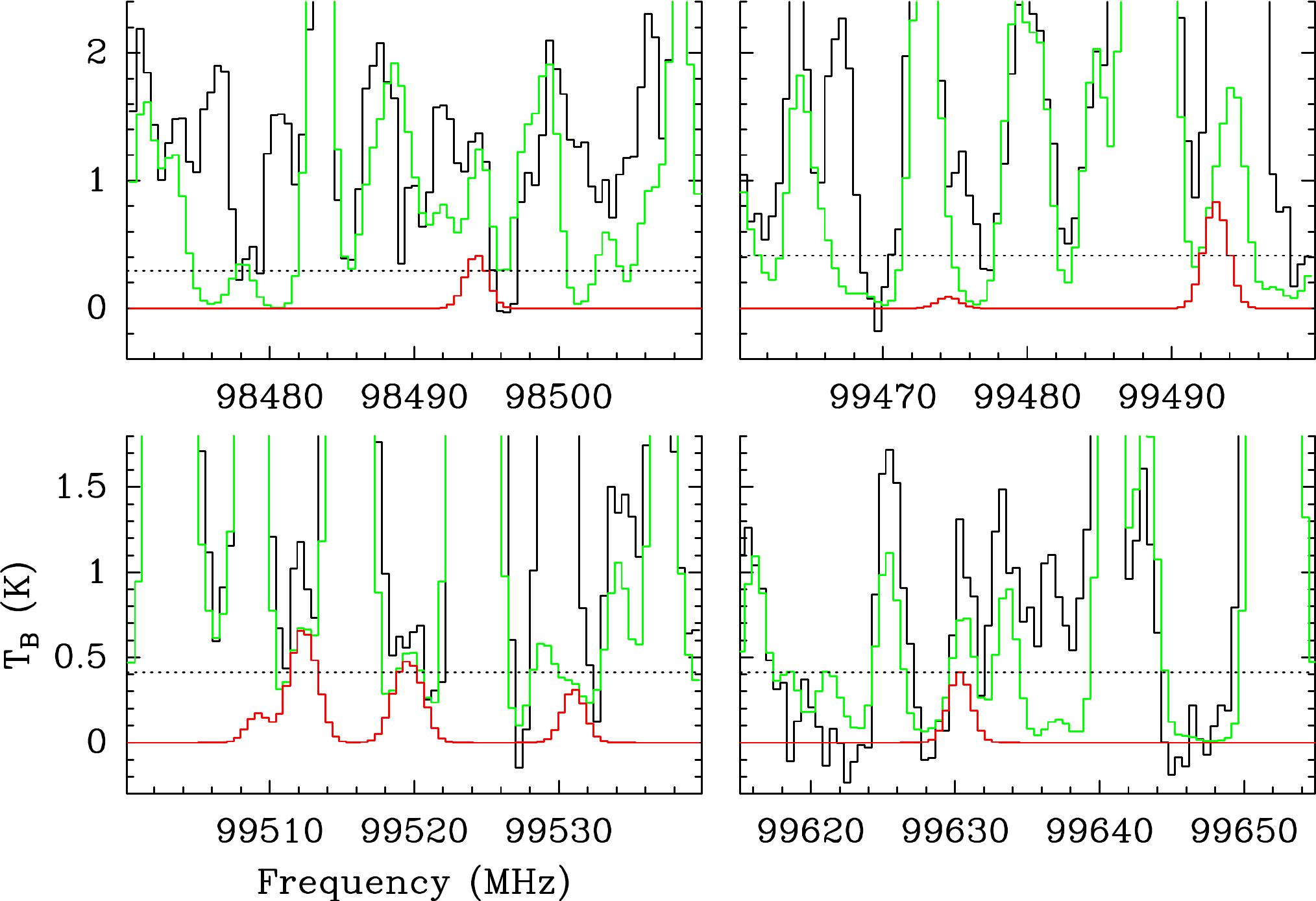}}}
\caption{EMoCA survey detail. Transitions of CH$_3$$^{34}$SH are shown, which are covered by the 
survey and that contribute significantly to the detected signal toward Sgr~B2(N2). The local 
thermodynamic equilibrium  synthetic spectrum of CH$_3$$^{34}$SH is displayed in red and 
overlaid on the observed spectrum shown in black. The green synthetic spectrum contains 
the contributions of all molecules identified in our survey so far, including the species 
shown in red. The y-axis is labeled in brightness temperature units. The dotted line indicates 
the $3\sigma$ noise level.
}
\label{f:spec_ch3sh_34s_ve0} 
\end{figure}

In order to search for rotational lines of the $^{34}$S isotopolog of methyl 
mercaptan in the EMoCA spectrum of Sgr~B2(N2), we started from the best-fit 
radiative-transfer model we obtained for the $^{32}$S isotopolog for this 
source \citep[][]{Holger2016}. We assumed the 
same size of the emitting region ($1.4\arcsec$), temperature (180~K), linewidth 
(5.4~km~s$^{-1}$), and centroid velocity (73.5~km~s$^{-1}$) as those derived for 
the main isotopolog. We followed \citet{1994ARA&A..32..191W} and assumed a $^{32}$S/$^{34}$S 
isotopic ratio of 22.5, equal to the Solar System value. Given the column 
density of $3.4\times10^{17}$~cm$^{-2}$ obtained for the main isotopolog,  this
implies a column density of $1.5\times10^{16}$~cm$^{-2}$ for the $^{34}$S 
isotopolog. With these parameters, we computed a synthetic spectrum of 
CH$_3$$^{34}$SH under the local thermodynamic equilibrium approximation 
using Weeds \citep[][]{refId0} and compare it to the observed EMoCA spectrum 
in Fig.~\ref{f:spec_ch3sh_34s_ve0}.

%%%%%%%%%%%%%%%%%%%%%%%%%%%%%%%%%%%%%%%%%%%%%%%%%%%%%%%%%%%%%%%%%%%%%%%%%%%%%%%%%%%%%%%%%
%%%%%%%%%%%%%%%%%%%%%%%%%%%%%%%%%%%%%%%%%%%%%%%%%%%%%%%%%%%%%%%%%%%%%%%%%%%%%%%%%%%%%%%%%
%%%%%%%%%%%%%%%%%%%%%%%%%%%%%%%%%%%%%%%%%%%%%%%%%%%%%%%%%%%%%%%%%%%%%%%%%%%%%%%%%%%%%%%%%
%%%%%%%%%%%%%%%%%%%%%%%%%%%%%%%%%%%%%%%%%%%%%%%%%%%%%%%%%%%%%%%%%%%%%%%%%%%%%%%%%%%%%%%%%

The synthetic spectrum of CH$_3$$^{34}$SH is consistent with the spectrum of
Sgr~B2(N2), but most of the strongest transitions expected for CH$_3$$^{34}$SH 
are, unfortunately, blended with emission from other molecules, which prevents 
their robust assignment to CH$_3$$^{34}$SH. Therefore, the molecule is not 
securely identified in the EMoCA spectrum. Still, the agreement between 
the observed and synthetic spectra for the line at 99519~MHz (which is 
contaminated by other species, though only in its wings) suggests that the molecule is
present at the level expected for the $^{32}$S/$^{34}$S isotopic ratio assumed
above.

%%%%%%%%%%%%%%%%%%%%%%%%%%%%%%%%%%%%%%%%%%%%%%%%%%%%%%%%%%%%%%%%%%%%%%%%%%%%%%%%%%%%%%%%%
%%%%%%%%%%%%%%%%%%%%%%%%%%%%%%%%%%%%%%%%%%%%%%%%%%%%%%%%%%%%%%%%%%%%%%%%%%%%%%%%%%%%%%%%%
\section{Conclusions}
\label{conclusion}
%%%%%%%%%%%%%%%%%%%%%%%%%%%%%%%%%%%%%%%%%%%%%%%%%%%%%%%%%%%%%%%%%%%%%%%%%%%%%%%%%%%%%%%%%
%%%%%%%%%%%%%%%%%%%%%%%%%%%%%%%%%%%%%%%%%%%%%%%%%%%%%%%%%%%%%%%%%%%%%%%%%%%%%%%%%%%%%%%%%

A new study of the rotational spectrum of the $^{34}$S isotopolog of CH$_3$SH 
was carried out in the frequency ranges of 49$-$510~GHz and 1.1$-$1.5~THz in order to provide 
accurate predictions for astronomical searches. The rotational transitions of the ground, the 
first, and the second excited torsional states were assigned up to high $J$ and $K_a$ quantum 
numbers (see Table~\ref{tbl:statisticInf}) and fit using a RAM Hamiltonian within experimental 
accuracy. The overall weighted rms deviation of the global fit of 3469 rotational transitions of 
the CH$_3$$^{34}$SH is 0.83, indicating that we have an appropriate set of parameters to provide 
reliable predictions to support astronomical observations.
CH$_3$$^{34}$SH was searched for with ALMA between 84~GHz and 114~GHz toward the hot molecular 
core Sgr B2(N2), but blends with emission lines of other species prevent its firm identification 
in this frequency range for this source. Nevertheless, the satisfactory agreement between 
observations and the astronomical model for the most uncontaminated CH$_3$$^{34}$SH line 
at 99519~MHz may make a secure detection possible in future observations for this source 
in a wider frequency range.

A calculated line list of this isotopic species, including information on 
intensities and calculated uncertainties, will be provided in the catalog 
section\footnote{https://cdms.astro.uni-koeln.de/classic/entries/} of the Cologne Database for 
Molecular Spectroscopy, CDMS \citep{ENDRES201695}. Files containing the experimental lines or 
parameters along with auxiliary files are available in the data section of the 
CDMS\footnote{https://cdms.astro.uni-koeln.de/classic/predictions/daten/Methanethiol/}.

%%%%%%%%%%%%%%%%%%%%%%%%%%%%%%%%%%%%%%%%%%%%%%%%%%%%%%%%%%%%%%%%%%%%%%%%%%%%%%%%%%%%%%%%%
%%%%%%%%%%%%%%%%%%%%%%%%%%%%%%%%%%%%%%%%%%%%%%%%%%%%%%%%%%%%%%%%%%%%%%%%%%%%%%%%%%%%%%%%%

\begin{acknowledgements}  
The present study was supported by the Deutsche Forschungsgemeinschaft (DFG) in the 
framework of the collaborative research grant SFB 956, project B3. O.Z. is funded by 
the DFG via the Ger{\"a}tezentrum "Cologne Center for Terahertz Spectroscopy".
The research in Kharkiv was carried out under support of the Volkswagen foundation. The 
assistance of Science and Technology Center in Ukraine is acknowledged (STCU partner project P686).
L.H.X. and R.M.L. received support from the Natural Sciences and Engineering Research Council 
of Canada.
\end{acknowledgements}

%%%%%%%%%%%%%%%%%%%%%%%%%%%%%%%%%%%%%%%%%%%%%%%%%%%%%%%%%%%%%%%%%%%%%%%%%%%%%%%%%%%%%%%%%
%%%%%%%%%%%%%%%%%%%%%%%%%%%%%%%%%%%%%%%%%%%%%%%%%%%%%%%%%%%%%%%%%%%%%%%%%%%%%%%%%%%%%%%%%
 
\bibliographystyle{aa} % style aa.bst
\bibliography{bibliography} % your references Yourfile.bib

%%%%%%%%%%%%%%%%%%%%%%%%%%%%%%%%%%%%%%%%%%%%%%%%%%%%%%%%%%%%%%%%%%%%%%%%%%%%%%%%%%%%%%%%%%
%%%%%%%%%%%%%%%%%%%%%%%%%%%%%%%%%%%%%%%%%%%%%%%%%%%%%%%%%%%%%%%%%%%%%%%%%%%%%%%%%%%%%%%%%%

%%%%%%%%%%%%%%%%%%%%%%%%%%%%%%%%%%%%%%%%%%%%%%%%%%%%%%%%%%%%%%%%%%%%%
%%%%%%%%%%%%%%%%%%%%%%%%%%%%%%%%%%%%%%%%%%%%%%%%%%%%%%%%%%%%%%%%%%%%%
%%%%%%   Table 2 + 3  %%%%%%%%%%%%%%%%%%%%%%%%%%%%%%%%%%%%%%%%%%%%%%%
%%%%%%%%%%%%%%%%%%%%%%%%%%%%%%%%%%%%%%%%%%%%%%%%%%%%%%%%%%%%%%%%%%%%%
%%%%%%%%%%%%%%%%%%%%%%%%%%%%%%%%%%%%%%%%%%%%%%%%%%%%%%%%%%%%%%%%%%%%%

\newpage
\longtab{
\begin{longtable}{llll}
\caption{\label{tbl:ParametersTable} List of parameters for CH$_3$$^{34}$SH}\\
\hline\hline
$n_{tr}$\textit{$^a$} & Operator\textit{$^b$} & Par.\textit{$^{c,d}$} & CH$_3$$^{34}$SH\textit{$^e$} \\
\hline
\endfirsthead
\caption{continued.}\\
\hline\hline
$n_{tr}$\textit{$^a$} & Operator\textit{$^b$} & Par.\textit{$^{c,d}$} & CH$_3$$^{34}$SH\textit{$^e$} \\
\hline
\endhead
\hline
    $2_{2,0}$ & $p_\alpha^2$                & $F$         & 15.018101(11)  \\
    $2_{2,0}$ & $(1-\cos 3\alpha)$          & $(1/2)V_3$  & 220.78189(36)  \\
    $2_{1,1}$ & $p_\alpha P_a$              & $\rho$      & 0.651352093(16)  \\
    $2_{0,2}$ & $P_a^2$                     & $A$  & 3.4254266(18)  \\
    $2_{0,2}$ & $P_b^2$                     & $B$  & 0.424734752(77)  \\
    $2_{0,2}$ & $P_c^2$                     & $C$  & 0.406539925(82)  \\
    $2_{0,2}$ & $(1/2)\{P_a{,}P_b\}$        & $2D_{ab}$   & $-0.0155173(17)$  \\
    $4_{4,0}$ & $(1-\cos 6\alpha)$          & $(1/2)V_6$ & $-0.28034(20)$  \\ 
    $4_{4,0}$ & $p_\alpha ^4$               & $F_m$      & $-0.11330(13)\times 10^{-2}$  \\
    $4_{3,1}$ & $p_\alpha ^3 P_a$           & $\rho_m$   & $-0.35743(34)\times 10^{-2}$  \\
    $4_{2,2}$ & $P^2(1-\cos 3\alpha)$       & $V_{3J}$ & $-0.20311808(74)\times 10^{-2}$ \\
    $4_{2,2}$ & $P_a^2(1-\cos 3\alpha)$     & $V_{3K}$ & $0.71456(12)\times 10^{-2}$  \\ 
    $4_{2,2}$ & $(P_b^2-P_c^2)(1-\cos 3\alpha)$ & $V_{3bc}$ & $-0.83920(12)\times 10^{-4}$  \\
    $4_{2,2}$ & $(1/2)\{P_a{,}P_b\}(1-\cos 3\alpha)$ & $V_{3ab}$ & $0.1207686(89)\times 10^{-1}$  \\  
    $4_{2,2}$ & $p^2_{\alpha} P^2$          & $F_J$ & $-0.3000547(70)\times 10^{-4}$  \\
    $4_{2,2}$ & $p^2_{\alpha} P_a^2$        & $F_K$ & $-0.47975(33)\times 10^{-2}$ \\  
    $4_{2,2}$ & $(1/2)\{P_a{,}P_c\}\sin 3\alpha$ & $D_{3ac}$ & $0.144733(33)\times 10^{-1}$ \\
    $4_{2,2}$ & $(1/2)\{P_b{,}P_c\}\sin 3\alpha$ & $D_{3bc}$ & $0.901353(76)\times 10^{-3}$ \\  
    $4_{1,3}$ & $p_\alpha P_aP^2$           & $\rho_J$ & $-0.413515(11)\times 10^{-4}$ \\
    $4_{1,3}$ & $p_\alpha P_a^3$            & $\rho_K$ & $-0.29544(14)\times 10^{-2}$ \\
    $4_{1,3}$ & $(1/2)\{P_a{,}(P_b^2-P_c^2)\}p_\alpha$ & $\rho_{bc}$ & $-0.19410(77)\times 10^{-4}$ \\
    $4_{0,4}$ & $P^4$                       & $-\Delta_J$ & $-0.523063(13)\times 10^{-6}$ \\
    $4_{0,4}$ & $P^2 P_a^2$                 & $-\Delta_{JK}$ & $-0.1737732(50)\times 10^{-4}$  \\
    $4_{0,4}$ & $P_a^4$                     & $-\Delta_K$ & $-0.69678(23)\times 10^{-3}$ \\
    $4_{0,4}$ & $P^2(P_b^2-P_c^2)$          & $-2\delta_J$ & $-0.435261(47)\times 10^{-7}$  \\
    $4_{0,4}$ & $(1/2)\{P_a^2{,}(P_b^2-P_c^2)\}$ &  $-2\delta_K$ & $-0.19133(74)\times 10^{-4}$ \\
    $4_{0,4}$ & $(1/2)P^2\{P_a{,}P_b\}$    & $D_{abJ}$ & $0.20067(82)\times 10^{-6}$  \\
    $6_{6,0}$ & $(1-\cos 9\alpha)$          & $(1/2)V_9$ & 0.11644(33) \\   
    $6_{6,0}$ & $p_\alpha ^6$               & $F_{mm}$ & $-0.3210(37)\times 10^{-5}$ \\
    $6_{5,1}$ & $p_\alpha ^5 P_a$           & $\rho_{mm}$ & $-0.1198(15)\times 10^{-4}$ \\
    $6_{4,2}$ & $P^2(1-\cos 6\alpha)$       & $V_{6J}$ & $-0.15588(23)\times 10^{-4}$ \\         
    $6_{4,2}$ & $P_a^2(1-\cos 6\alpha)$     & $V_{6K}$ & $-0.846(28)\times 10^{-4}$ \\ 
    $6_{4,2}$ & $(1/2)\{P_a{,}P_b\}(1-\cos 6\alpha)$ &  $V_{6ab}$ & $0.7833(69)\times 10^{-4}$ \\ 
    $6_{4,2}$ & $(P_b^2-P_c^2)(1-\cos 6\alpha)$ & $V_{6bc}$ & $-0.26332(24)\times 10^{-4}$ \\
    $6_{4,2}$ & $(1/2)\{P_a{,}P_c\}\sin 6\alpha$  & $D_{6ac}$ & $0.2798(22)\times 10^{-3}$ \\     
    $6_{4,2}$ & $(1/2)\{P_b{,}P_c\}\sin 6\alpha$ & $D_{6bc}$ & $0.1747(11)\times 10^{-4}$ \\  
    $6_{4,2}$ & $(1/2)\{P_b{,}P_c{,}p_\alpha ^2{,}\sin 3\alpha\}$ & $D_{3bcm}$ & $0.5133(20)\times 10^{-5}$ \\
    $6_{4,2}$ & $p_\alpha ^4P^2$            & $F_{mJ}$ & $0.2179(37)\times 10^{-8}$ \\    
    $6_{4,2}$ & $p_\alpha ^4P_a^2$          & $F_{mK}$ & $-0.1827(24)\times 10^{-4}$ \\ 
    $6_{3,3}$ & $p_\alpha ^3P_aP^2$         & $\rho_{mJ}$ & $0.502(10)\times 10^{-8}$ \\   
    $6_{3,3}$ & $p_\alpha ^3P_a^3$          & $\rho_{mK}$ & $-0.1448(21)\times 10^{-4}$ \\               
    $6_{3,3}$ & $(1/2)\{P_a{,}P_b{,}P_c{,}p_\alpha{,}\sin 3\alpha\}$ & $\rho_{bc3}$ & $0.3849(14)\times 10^{-5}$ \\
    $6_{2,4}$ & $P^4(1-\cos 3\alpha)$       & $V_{3JJ}$ & $0.4541(19)\times 10^{-8}$ \\
    $6_{2,4}$ & $P^2P_a^2(1-\cos 3\alpha)$   & $V_{3JK}$ & $-0.22294(16)\times 10^{-6}$ \\
    $6_{2,4}$ & $P_a^4(1-\cos 3\alpha)$     & $V_{3KK}$ & $0.4741(21)\times 10^{-6}$ \\
    $6_{2,4}$ & $(1/2)P^2\{P_a{,}P_b\}(1-\cos 3\alpha)$ & $V_{3abJ}$ & $-0.15652(65)\times 10^{-6}$ \\  
    $6_{2,4}$ & $P^2(P_b^2-P_c^2)(1-\cos 3\alpha)$ & $V_{3bcJ}$ & $0.7677(48)\times 10^{-9}$ \\       
    $6_{2,4}$ & $(1/2)\{P_b^2{,}P_c^2\}\cos 3\alpha$ & $V_{3b2c2}$ & $0.400(14)\times 10^{-8}$  \\  
    $6_{2,4}$ & $p^2_{\alpha} P^4$          & $F_{JJ}$ & $0.1207(22)\times 10^{-9}$ \\       
    $6_{2,4}$ & $p^2_{\alpha} P_a^2 P^2$    & $F_{JK}$ & $0.614(12)\times 10^{-8}$ \\       
    $6_{2,4}$ & $p^2_{\alpha} P_a^4$        & $F_{KK}$ & $-0.617(10)\times 10^{-5}$ \\ 
    $6_{2,4}$ & $(1/2)\{P_b^2{,}P_c^2\}p^2_{\alpha}$ & $F_{b2c2}$ & $-0.338(17)\times 10^{-9}$ \\
    $6_{2,4}$ & $(1/2)P^2\{P_a{,}P_c\}\sin 3\alpha$ & $D_{3acJ}$ & $-0.512(16)\times 10^{-7}$  \\  
    $6_{2,4}$ & $(1/2)\{P_a^3{,}P_c\}\sin 3\alpha$ & $D_{3acK}$ & $-0.601(26)\times 10^{-6}$  \\  
    $6_{2,4}$ & $(1/2)P^2\{P_b{,}P_c\}\sin 3\alpha$ & $D_{3bcJ}$ & $-0.12267(38)\times 10^{-7}$  \\  
    $6_{1,5}$ & $p_\alpha P_aP^4$           & $\rho_{JJ}$ & $0.19908(96)\times 10^{-9}$ \\
    $6_{1,5}$ & $p_\alpha P_a^3 P^2$        & $\rho_{JK}$ & $0.4053(64)\times 10^{-8}$ \\
    $6_{1,5}$ & $p_\alpha P_a^5$            & $\rho_{KK}$ & $-0.1289(26)\times 10^{-5}$ \\
    $6_{1,5}$ & $(1/2)P^2\{P_a{,}(P_b^2-P_c^2)\}p_\alpha$ & $\rho_{bcJ}$ & $0.934(23)\times 10^{-10}$ \\
    $6_{1,5}$ & $(1/2)\{P_a{,}P_b^2{,}P_c^2\}p_\alpha$ & $\rho_{b2c2}$ & $-0.6003(67)\times 10^{-9}$ \\
    $6_{0,6}$ & $P^6$                       & $\Phi_J$ & $-0.26953(51)\times 10^{-12}$ \\
    $6_{0,6}$ & $P^4 P_a^2$                 & $\Phi_{JK}$ & $0.6360(20)\times 10^{-10}$ \\
    $6_{0,6}$ & $P^2P_a^4$                  & $\Phi_{KJ}$ & $0.1146(14)\times 10^{-8}$  \\    
    $6_{0,6}$ & $P_a^6$                     & $\Phi_K$ & $-0.913(29)\times 10^{-7}$ \\
    $6_{0,6}$ & $(1/2)P^2\{P_a^2{,}(P_b^2-P_c^2)\}$ & $2\phi_{JK}$ & $0.887(21)\times 10^{-10}$ \\
    $8_{6,2}$ & $P^2(1-\cos 9\alpha)$       & $V_{9J}$ & $-0.2513(38)\times 10^{-5}$ \\  
    $8_{6,2}$ & $P_a^2(1-\cos 9\alpha)$      & $V_{9K}$ & $-0.548(35)\times 10^{-4}$  \\  
    $8_{6,2}$ & $(1/2)\{P_a{,}P_c\}\sin 9\alpha$ & $D_{9ac}$ & $0.648(14)\times 10^{-4}$ \\
    $8_{4,4}$ & $P^4(1-\cos 6\alpha)$       & $V_{6JJ}$ & $0.1254(42)\times 10^{-8}$ \\     
    $8_{4,4}$ & $P^2 P_a^2(1-\cos 6\alpha)$ & $V_{6JK}$ & $-0.2418(23)\times 10^{-7}$ \\     
    $8_{4,4}$ & $P^2(P_b^2-P_c^2)(1-\cos 6\alpha)$ & $V_{6bcJ}$ & $0.695(12)\times 10^{-9}$ \\       
    $8_{4,4}$ & $(1/2)\{P_b^2{,}P_c^2\}\cos 6\alpha$ & $V_{6b2c2}$ & $0.674(32)\times 10^{-8}$   \\

\hline
\end{longtable}
\tablefoot{$^{a}$ $n=t+r$, where $n$ is the total order of the operator, $t$ is the order of the torsional part, 
and $r$ is the order of the rotational part, respectively. 
The ordering scheme of \citet{NAKAGAWA:1987} is used. $^{b}$ \{A,B,C,D\} = ABCD + DCBA. \{A,B,C\} = ABC + CBA. \{A,B\} = AB + BA. 
The product of the operator in the first column of a given row and the parameter in the third column of that row gives the term 
currently used in the torsion-rotation Hamiltonian of the program, except for $F$, $\rho$, and $A_{\rm RAM}$, 
which occur in the Hamiltonian in the form of $F(p_a + \rho P_a)^2 + A_{\rm RAM}P_a^2$. 
$^{c}$ Parameter nomenclature based on the subscript procedure of \citet{XU:2008305}. 
$^{d}$ Values of the parameters are in cm$^{-1}$, except for $\rho$, which is unitless. 
$^{e}$ Statistical uncertainties are given in parentheses as one standard uncertainty in units of the last digits. }
}

%%%%%%%%%%%%%%%%%%%%%%%%%%%%%%%%%%%%%%%%%%%%%%%%%%%%%%%%%%%%%%%%%%%%%%%%%%%%%%%%%%%%%%%%%%
%%%%%%%%%%%%%%%%%%%%%%%%%%%%%%%%%%%%%%%%%%%%%%%%%%%%%%%%%%%%%%%%%%%%%%%%%%%%%%%%%%%%%%%%%%

\newpage
\longtab{
\begin{longtable}{llllll}
\caption{\label{tbl:ParametersComparison} Comparison of the low-order parameters of CH$_3$$^{34}$SH, CH$_3$$^{32}$SH, and CH$_3$$^{32}$SD}\\
\hline\hline
$n_{tr}$\textit{$^a$} & Operator\textit{$^b$} & Par.\textit{$^{c,d}$} & CH$_3$$^{34}$SH\textit{$^e$} & CH$_3$$^{32}$SH\textit{$^{e,f}$} & CH$_3$$^{32}$SD\textit{$^{e,f}$}\\
\hline
\endfirsthead
\caption{continued.}\\
\hline\hline
$n_{tr}$\textit{$^a$} & Operator\textit{$^b$} & Par.\textit{$^{c,d}$} & CH$_3$$^{34}$SH\textit{$^e$} & CH$_3$$^{32}$SH\textit{$^{e,f}$} & CH$_3$$^{32}$SD\textit{$^{e,g}$} \\
\hline
\endhead
\hline
    $2_{2,0}$ & $p_\alpha^2$               & $F$ & 15.018101(11) & 15.04020465(66) & 10.3520639(31)\\
    $2_{2,0}$ & $(1/2)(1-\cos 3\alpha)$    & $V_3$ & 441.56378(72) & 441.442236(10) & 435.42500(24)\\
    $2_{1,1}$ & $p_\alpha P_a$             & $\rho$ & 0.651352093(16) & 0.651856026(13) & 0.493517098(11) \\
    $2_{0,2}$ & $P_a^2$                    & $A$ & 3.4254266(18) & 3.42808445(84) & 2.59513758(20) \\
    $2_{0,2}$ & $P_b^2$                    & $B$ & 0.424734752(77) & 0.43201954(87) & 0.42517153(13) \\
    $2_{0,2}$ & $P_c^2$                    & $C$ & 0.406539925(82) & 0.41325076(83) & 0.39176839(13) \\
    $2_{0,2}$ & $\{P_a{,}P_b\}$            & $D_{ab}$ & $-0.00775864(85)$ & $-0.0073126(59)$ & 0.0053655(12)\\
    $4_{4,0}$ & $(1/2)(1-\cos 6\alpha)$    & $V_6$ & $-0.56068(40)$ & $-0.572786(15)$ & $-0.86918(20)$ \\ 
    $4_{4,0}$ & $p_\alpha ^4$              & $F_m$ & $-0.11330(13)\times 10^{-2}$ & $-0.114016(10)\times 10^{-2}$ & $-0.38535(20)\times 10^{-3}$ \\
    $4_{3,1}$ & $p_\alpha ^3 P_a$          & $\rho_m$ & $-0.35743(34)\times 10^{-2}$ & $-0.360009(28)\times 10^{-2}$ & $-0.93969(42)\times 10^{-3}$ \\
    $4_{2,2}$ & $P^2(1-\cos 3\alpha)$      & $V_{3J}$ & $-0.20311808(74)\times 10^{-2}$ & $-0.217540(84)\times 10^{-2}$ & $-0.19405784(64)\times 10^{-2}$ \\
    $4_{2,2}$ & $P_a^2 (1-\cos 3\alpha)$   & $V_{3K}$ & $0.71456(12)\times 10^{-2}$ & $0.724978(19)\times 10^{-2}$ & $0.685147(15)\times 10^{-2}$ \\ 
    $4_{2,2}$ & $(P_b^2-P_c^2)(1-\cos 3\alpha)$ & $V_{3bc}$ & $-0.83920(12)\times 10^{-4}$ & $-0.92104(47)\times 10^{-4}$ & $-0.15679(13)\times 10^{-3}$ \\
    $4_{2,2}$ & $\{P_a{,}P_b\}(1-\cos 3\alpha)$ & $V_{3ab}$ & $0.603843(45)\times 10^{-2}$ & $0.61562(30)\times 10^{-2}$ & $0.443895(41)\times 10^{-2}$ \\  
    $4_{2,2}$ & $p^2_{\alpha} P^2$         & $F_J$ & $-0.3000547(70)\times 10^{-4}$ & $-0.8106(38)\times 10^{-4}$ & $-0.2823067(53)\times 10^{-4}$ \\
    $4_{2,2}$ & $p^2_{\alpha} P_a^2$       & $F_K$ & $-0.47975(33)\times 10^{-2}$ & $-0.483287(30)\times 10^{-2}$ & $-0.123322(32)\times 10^{-2}$ \\  
    $4_{2,2}$ & $p_\alpha ^2\{P_a{,}P_b\}$ & $F_{ab}$ & $ - $ & $0.843(45)\times 10^{-4}$ & $0.7807(70)\times 10^{-4}$ \\
    $4_{2,2}$ & $2p^2_{\alpha}(P_b^2-P_c^2)$ & $F_{bc}$ & $ - $ & $0.0536(41)\times 10^{-4}$ & $0.102769(43)\times 10^{-4}$ \\
    $4_{2,2}$ & $\{P_a{,}P_c\}\sin 3\alpha$ & $D_{3ac}$ & $0.072366(16)\times 10^{-1}$ & $0.1036(15)\times 10^{-1}$ & $0.11336(45)\times 10^{-1}$ \\
    $4_{2,2}$ & $\{P_b{,}P_c\}\sin 3\alpha$ & $D_{3bc}$ & $0.450676(38)\times 10^{-3}$ & $0.665(14)\times 10^{-3}$ & $0.140496(27)\times 10^{-2}$ \\  
    $4_{1,3}$ & $p_\alpha P_a P^2$         & $\rho_J$ & $-0.413515(11)\times 10^{-4}$ & $-0.4726(54)\times 10^{-4}$ & $-0.369536(28)\times 10^{-4}$ \\
    $4_{1,3}$ & $p_\alpha P_a^3$           & $\rho_K$ & $-0.29544(14)\times 10^{-2}$ & $-0.30381(74)\times 10^{-2}$  & $-0.75792(11)\times 10^{-3}$\\
    $4_{1,3}$ & $p_\alpha \{P_a^2{,}P_b\}$ & $\rho_{ab}$ & $ - $ & $0.999(67)\times 10^{-4}$ & $0.1017(10)\times 10^{-3}$ \\  
    $4_{1,3}$ & $p_\alpha \{P_a{,}(P_b^2-P_c^2)\}$ & $\rho_{bc}$ & $-0.09705(39)\times 10^{-4}$ & $-0.0462(39)\times 10^{-4}$ & $ - $  \\
    $4_{0,4}$ & $-P^4$                     & $\Delta_J$ & $0.523063(13)\times 10^{-6}$ & $0.538140(23)\times 10^{-6}$ & $0.4876778(87)\times 10^{-6}$ \\
    $4_{0,4}$ & $-P^2 P_a^2$               & $\Delta_{JK}$ & $0.1737732(50)\times 10^{-4}$ & $-0.066(26)\times 10^{-5}$ & $0.143777(64)\times 10^{-4}$ \\
    $4_{0,4}$ & $-P_a^4$                   & $\Delta_K$ & $0.69678(23)\times 10^{-3}$ & $0.7425(48)\times 10^{-3}$ & $0.178662(17)\times 10^{-3}$ \\
    $4_{0,4}$ & $-2P^2(P_b^2-P_c^2)$       & $\delta_J$ & $0.217630(24)\times 10^{-7}$ & $0.224788(88)\times 10^{-7}$ & $0.384636(10)\times 10^{-7}$ \\
    $4_{0,4}$ & $-\{P_a^2{,}(P_b^2-P_c^2)\}$ & $\delta_K$ & $0.09566(37)\times 10^{-4}$ & $0.10483(32)\times 10^{-4}$ & $0.090811(35)\times 10^{-4}$ \\
    $4_{0,4}$ & $P^2\{P_a{,}P_b\}$         & $D_{abJ}$ & $0.10033(41)\times 10^{-6}$ & $-0.956(60)\times 10^{-7}$ & $ - $ \\    
    $4_{0,4}$ & $\{P_a^3{,}P_b\}$          & $D_{abK}$ & $ - $ & $0.202(23)\times 10^{-4}$ & $0.2750(34)\times 10^{-4}$ \\    
     &                                     & $\theta_{\rm RAM}$ & $-$0.15$^\circ$ & $-$0.14$^\circ$ & 0.14$^\circ$  \\      
\hline
\end{longtable}
\tablefoot{$^{a}$ $n=t+r$, where $n$ is the total order of the operator, $t$ is the order of the torsional part, and $r$ 
is the order of the rotational part, respectively. The ordering scheme of \citet{NAKAGAWA:1987} is used. $^{b}$ \{A,B,C,D\} = ABCD + DCBA. 
\{A,B,C\} = ABC + CBA. \{A,B\} = AB + BA. The product of the operator in the first column of a given row and the parameter in the third column 
of that row gives the term currently used in the torsion-rotation Hamiltonian of the program, except for \textit{F}, $\rho$ and $A_{\rm RAM}$, 
which occur in the Hamiltonian in the form of $F(p_a + \rho P_a)^2 + A_{\rm RAM}P_a^2$. $^{c}$ Parameter nomenclature is based on the subscript procedure 
of \citet{XU:2008305}. $^{d}$ Values of the parameters are in cm$^{-1}$, except for $\rho$, which is unitless, and except for \textit{$\theta_{\rm RAM},$} which is in degrees. 
$^{e}$ Statistical uncertainties are given in parentheses as one standard uncertainty in units of the last digits. 
$^{f}$ Not all the parameters used for the analysis in \citet{Xu:2012} and \citet{Zakharenko:2019SD} listed here.}
}

\end{document}